\documentclass[12pt]{article}
\usepackage[twoside,pdftex,papersize={210mm,280mm}, total={16.8cm,23.6cm}, left=2cm, top=2cm, right=2cm, bottom=2cm, includehead=true]{geometry}
\usepackage{amssymb}
\usepackage[affil-it]{authblk}
\usepackage[fleqn]{amsmath}
\usepackage{lineno}
\usepackage{epsfig}
\usepackage{subfigure}
\usepackage{hyperref}
\usepackage{cancel}

\begin{document}

\title{Estimates of $Z$ Boson and $J/\psi$ production cross sections at the LHC}

\author{T.~Alexopoulos$^{1}$,  S.~Leontsinis$^{1,2}$\footnote{stefanos.leontsinis@cern.ch}}

\affil{$^1$National Technical University of Athens}
\affil{$^2$Brookhaven National Laboratory}

\maketitle

\begin{abstract}
We calculate the leading order cross section for the associated production of $Z$ and $J/\psi$. Processes that include associated production of electroweak bosons and heavy quarlonium can give valuable insight to the production mechanism of quarkonia. We conclude that this process is accessible by the LHC statistics.
\end{abstract}

\maketitle

\section{Introduction}

The quarkonium is a conceptually simple system that can be described by quantum chromodynamics (QCD) as a pair of quarks with the same flavour. However, the formation and kinematics of heavy quarkonia is still under investigation with various models (perturbative and non-perturbative QCD) trying to explain it \cite{quarkonium_group}.

Theoretical progress has been made on the factorization between the short distance physics of the heavy-quark creation and the long distance physics of the bound state formation. The effective field theory, including the colour-octet mechanisms, based on non-relativistic QCD (NRQCD) \cite{ref:NRQCD} replaced the colour singlet model \cite{ref:CS1,ref:CS2}. The colour-octet mechanism was used to cancel the infrared divergences in the decay widths of $P$-wave and $D$-wave heavy quarkonia \cite{ref:Pwave, ref:Dwave}. Another very interesting model under test is the colour evaporation model \cite{ref:CE1, ref:CE2}.

The colour singlet model requires that the type of quarkonia produced is determined by the state of the original quarks and that is what gives the name to the model. The colour evaporation model doesn't demand the quark pair to be produced in a colour singlet state. It can be produced as colour octet state and the colour and spin is then modified via soft interactions with the colour field. The colour octet mechanism proposes that the quark pairs produced by the hard process are not produced with the quantum numbers of the physical quarkonia but evolve into the quarkonia state through radiation of soft gluons.

The framework of NRQCD postulates that colour-octet processes associated with Fock state components of the quarkonia contribute to the cross section. The difference with the colour-singlet model is that the quarkonia system is produced from heavy quark pairs generated at short distances in colour-octet states with the emission of soft gluons (when the quark pair has the quarkonia size), while colour-singlet implies that the quark pair is generated with quantum numbers of spin and angular momentum of the meson. In NRQCD high energy scales (of the order of $m_Q$) are separated from low energy scales in quarkonia production or annihilation rates. The NRQCD Lagrangian is derived from the QCD Lagrangian by integrating out energy-momentum modes of order of $m_Q$ and higher \cite{ref:NRQCD_lagrangian}. It has the form

\begin{equation}
\mathcal{L}=\mathcal{L_{\mathrm{heavy}}}+\mathcal{L}_{\mathrm{light}}+\delta\mathcal{L}\ ,
\end{equation} 

\noindent where $\mathcal{L}_{\mathrm{light}}$ is the standard relativistic QCD Lagrangian for gluons and light quarks 

\begin{equation}
\mathcal{L}_{\mathrm{light}}=-\frac{1}{2}\mathrm{Tr}(G_{\mu\nu}G^{\mu\nu})+\sum\bar{q}i\cancel{D}q\ 
\end{equation}

\noindent and $\mathcal{L}_\mathrm{heavy}$ describes the low-momentum modes associated with the heavy quarks

\begin{equation}
\mathcal{L_{\mathrm{heavy}}}=\psi^\dagger\left(iD_0+\frac{\mathbf{D^2}}{2m_{Q}}\right)\psi+\chi^\dagger\left(iD_0+\frac{\mathbf{D^2}}{2m_{Q}}\right)\chi\ .
\end{equation}

We denote with $D^\mu$ the covariant derivate, which is given by $D^\mu = \partial^\mu + i g_s A^\mu$, where the SU(3) gauge field is $A^\mu=(\phi, \mathbf A^\mu)$ and $g_s$ is the QCD coupling given by $\sqrt{4 \pi\alpha_s}$. $\psi$ is the Pauli spinor field that annihilates a heavy quark and $\chi$ is the Pauli spinor field that creates the heavy anti-quark. $D_0$ and $\mathbf{D}$ are the time and space components of $D^\mu$. The correction term $\delta\mathcal{L}$ includes all possible operators consistent with QCD symmetries.

The measurement of the cross section of the associated production of electroweak bosons and heavy quarkonia cross section is crucial because it is going to shed light on the production of the quarkonia formation. Also an excess of events might be a signal for decays of a fermiophobic Higgs boson \cite{ref:fermiophobic}.

\begin{figure}[h!]
	\begin{center}
		\includegraphics[scale=0.85]{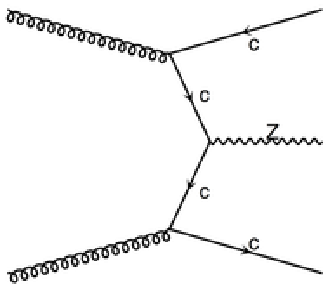}
		\includegraphics[scale=0.85]{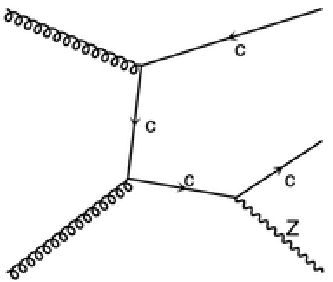}
		\includegraphics[scale=0.85]{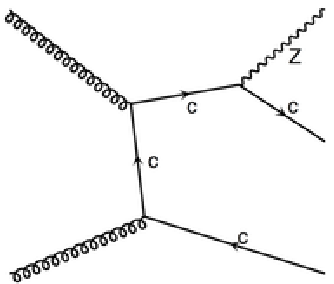}
		\includegraphics[scale=0.85]{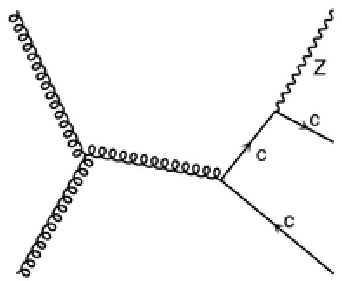}
		\includegraphics[scale=0.85]{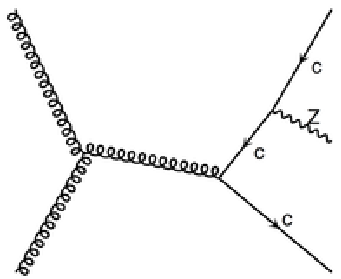}\\
		\includegraphics[scale=0.85]{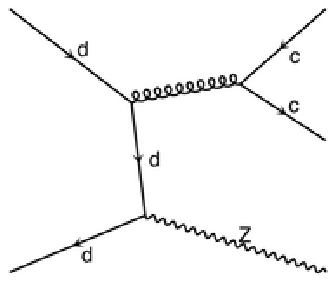}
		\includegraphics[scale=0.85]{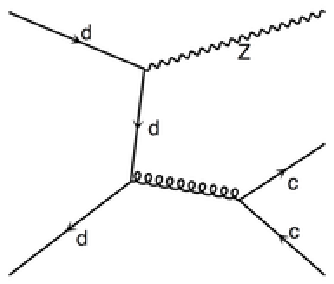}
		\includegraphics[scale=0.85]{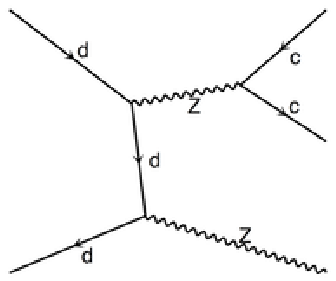}
		\includegraphics[scale=0.85]{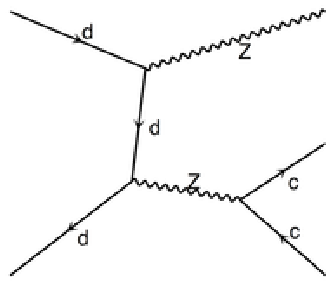}
		\includegraphics[scale=0.85]{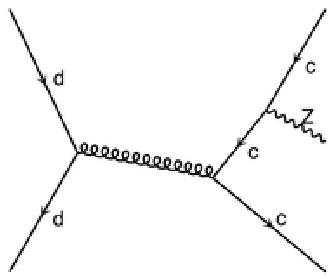}
		\caption{Leading order Feynman diagrams for the associated production of a $Z$ boson and a $J/\psi$.}
 		\label{fig:feyn_diagrams}
	\end{center}
\end{figure}

It is our purpose to estimate the production cross section for the processes $q+\bar{q}\to Z+c\bar{c}\,(^{2S+1}L_J^{(n)})$, where $q=u,d,c,s$ and $g+g\to Z+c\bar{c}\,(^{2S+1}L_J^{(n)})$, where $S$ is the definite spin, $L$ the orbital angular momentum, $J$ the total angular momentum and $c$ the colour multiplicity, considering $c=1$ (colour singlet) and $c=8$ (colour-octet). We will focus on colour-octet states, in which the quarkonia system is produced in short distances and evolve into colour singlet state by the emission of soft gluons. Examples of tree-level Feynman diagrams can be seen in Figure \ref{fig:feyn_diagrams}.

The reason we consider these particular processes, is because they are experimentally favored, due to clean signals from the purely leptonic decays of the $Z\to l^+l^-,\, W\to l\nu$ and the quarkonia ($\Upsilon\to\mu^+\mu^-,\, J/\psi\to\mu^+\mu^-$) with the highly suppressed background. Searches have been performed in the past at CDF ($\Upsilon+W^{\pm}$ or $Z$) with no indication of signal \cite{ref:CDFZUpsilon}, but recently ATLAS at CERN had an observation of $W^\pm+J/\psi$ with a $5.3\,\sigma$ significance at $\sqrt{s}=7\,\mathrm{TeV}$ \cite{ref:ATLASWJpsi}. Unfortunately the theoretical predictions are not in good agreement with the experimental measurements. This discrepancy may be resolved by including contributions from higher-order diagrams and intrinsic charm content in the proton \cite{intrinsiccharm}.

Our results were obtained using MADONIA (based on MadGraph \cite{MADONIA} matrix element generator) which allows the calculation of the cross sections of these processes. MadGraph provides partonic helicity amplitudes for tree-level Standard Model processes. We focus on the result and the possibility that these processes are accessible by the LHC, rather than a detailed numerical analysis. We show that the process of $Z+J/\psi$ is also accessible with the current statistics of the LHC.

\section{Cross Section Results}
There are many studies in the literature involving the associated production of electroweak bosons ($W$ or $Z$) and heavy quarkonia ($\Upsilon$ or $J/\psi$) \cite{QW1,QW2,QW3,QW4} and many for the specific process of $Z+J/\psi$ that we study in this note \cite{ZJpsi1, ZJpsi2, ZJpsi3}.

We consider two processes that contribute to the $p+p\to J/\psi+Z$ at leading order. First we consider $g+g\to c\bar{c}[n]+Z$ and then $q+\bar{q}\to c\bar{c}[n]+Z$, where $q$ can be either $u,d,s,c$ and $n=^3 \mathrm{S}_1^{(8)},\ ^1\mathrm{S}_0^{(8)}$ or $^3\mathrm{S}_1^{(1)}$.

The cross section of the associated production of $J/\psi$ and $Z$ in the framework of NRQCD is given by

\begin{equation}
\sigma(pp\to\mathcal{Q}+Z+X)=\sum_{n}\hat{\sigma}(pp\to c\bar{c}(n)+Z+X)\langle\mathcal{O}^{\mathcal{Q}}(n)\rangle\ ,
\end{equation}

\noindent where $\hat{\sigma}(pp\to c\bar{c}(n)+Z+X)$ is the short distance cross section and $\langle\mathcal{O}^{\mathcal{Q}}(n)\rangle$ the long distance matrix element (LDME). Effects of the order of $Q^2/m^2_Q\geq 1$ ($m_Q$ being the quark mass and $Q$ the momentum transfer in a production process), that are encoded in short distance coefficients, can be estimated using pertubation theory. On the other hand, effects of the order of $Q^2/m_Q^2<1$ hadronization, are factorized into long distance matrix elements, expressed in powers of $\upsilon$ and measured from lattice simulations or from experimental data. LDME are expected to be process-independent, not to depend on the production mechanism of the perturbative heavy quarks and at present they can't be computed from first principles.

The LDME are related to the non-perturbative transition probabilities from a $Q\bar{Q}$ system in a quarkonia state and they scale with a definite power of the intrinsic heavy quark velocity $\upsilon$. Thus, studies including $\Upsilon$ may be more suitable for the understanding of the NRQCD factorization formalism, since the mass of the bottomonium is heavier than the charmonium of the order of about $3$, implying smaller $\upsilon^2$, thus faster convergence\footnote{Charmonium ground state: $\upsilon^2\sim 0.3$, bottomonium ground state: $\upsilon^2\sim 0.1$.}. In addition, the asymptotic behavior for the $\Upsilon$ is reached at much higher values of transverse momentum ($p_{\mathrm{T}}$), due to the fact that $m_b>m_c$.

Charmonium on the other hand has the advantage that its mass is closer to $\Lambda_{\mathrm{QCD}}$ than the bottomonium. This enables us to perform a non-relativistic treatment of a quarkonia system for the understanding of the production and decay of bound states of heavy quarks. This strategy makes it possible to embed the present	 approach in the framework of NRQCD.

The parameter values used as an input for the calculations are \cite{ref:PDG}:
\begin{itemize}
\item CTEQ6L1 PDF set
\item $m_{Z}=91.18\,\mathrm{GeV}$
\item $\alpha_S(m_Z)=0.1184$
\item $m_{c}=1.275\,\mathrm{GeV},\ m_{u}=2.3\,\mathrm{MeV},\ m_{d}=4.8\,\mathrm{MeV}, \ m_{s}=95.5\,\mathrm{MeV}$
\item $\alpha=7.297\times 10^{-3}$
\item $\mu_R=\mu_F=m_Z$
\item NRQCD matrix elements for the charmonium production \cite{NRQCDitems}
\begin{itemize}
	\item $\langle\mathcal{O}(J/\psi)[^3S_1^{(1)}]\rangle=1.64\,\mathrm{GeV}^3$
	\item $\langle\mathcal{O}(J/\psi)[^3S_1^{(8)}]\rangle=0.3\times 10^{-3}\,\mathrm{GeV}^3$
	\item $\langle\mathcal{O}(J/\psi)[^1S_0^{(8)}]\rangle=8.9\times 10^{-2}\,\mathrm{GeV}^3$
\end{itemize}
\end{itemize}

Additional kinematic cuts that were applied to the $J/\psi$, following the acceptance of the experiments in LHC. The transverse momentum of the $J/\psi$ is required to be $p_\mathrm{T}^{J/\psi}>8\,\mathrm{GeV}$ and its rapidity $|y^{J/\psi}|<2.4$.

The results of every process is presented in table \ref{tab:numerical_results_analytical}, where only statistical errors are shown. Processes of the format $q+\bar{q}\to c\bar{c}[^{3}S_1^{(1)}]+Z$ are expected to have very low cross sections. This is because the $c$-quark line of the charmonium is connected with the $q$-quark line by the gluon that transmits colour to the $c\bar{c}$. This was checked with our simulation and can be seen from the absence of these processes from table \ref{tab:numerical_results_analytical} and the very low cross section of the process where $q=c$.

\begin{table*}[h]
\begin{center}\small
\begin{tabular}{ | c | c | c | c | c | c | }
  \hline
  & \multicolumn{3}{|c|}{cross section [$\mathrm{fb}$]} \\
  \hline
  Process & $\sqrt{s}=7\,\mathrm{TeV}$ & $\sqrt{s}=8\,\mathrm{TeV}$ & $\sqrt{s}=14\,\mathrm{TeV}$\\\hline\hline
  $g+g\to Z + c\bar{c} [^3\mathrm{S}_1^{(8)}]$       & $11.3   \pm 3.6 $ & $ 14.1  \pm 5.0  $&  $ 32.8  \pm 12.1$\\\hline
  $c+\bar{c}\to Z + c\bar{c} [^3\mathrm{S}_1^{(8)}]$ & $15.7   \pm 5.2 $ & $ 19.7  \pm 6.0  $&  $ 47.4  \pm 26.1$\\\hline
  $u+\bar{u}\to Z + c\bar{c} [^3\mathrm{S}_1^{(8)}]$ & $195.5  \pm 20.4$ & $ 204.4 \pm 29.7 $&  $ 408.3 \pm 50.7$\\\hline
  $d+\bar{d}\to Z + c\bar{c} [^3\mathrm{S}_1^{(8)}]$ & $148.0  \pm 21.3$ & $ 157.4 \pm 19.8 $&  $ 342.5 \pm 40.4$\\\hline
  $s+\bar{s}\to Z + c\bar{c} [^3\mathrm{S}_1^{(8)}]$ & $56.0   \pm 10.7$ & $ 70.3  \pm 13.3 $&  $ 181.1 \pm 54.4$\\\hline
  \hline
  $g+g\to Z + c\bar{c} [^1\mathrm{S}_0^{(8)}]$       & $281.0 \pm 36.0$ & $300.5 \pm 42.5 $ & $823.1 \pm 101.3$\\\hline
  $c+\bar{c}\to Z + c\bar{c} [^1\mathrm{S}_0^{(8)}]$ & $0.4   \pm 2.9 $ & $1.1   \pm 1.7  $ & $8.2 \pm 8.4$\\\hline
  \hline
  $g+g\to Z + c\bar{c} [^3\mathrm{S}_1^{(1)}]$       & $7.0 \pm 0.9$ & $9.1 \pm 1.0$&  $20.5 \pm 2.7$\\\hline
  $c+\bar{c}\to Z + c\bar{c} [^3\mathrm{S}_1^{(1)}]$ & $1.8 \pm 0.4$ & $2.1 \pm 0.8$&  $ 5.4 \pm 1.9$\\\hline
  \hline
\end{tabular}
\caption{Cross sections tree level at $\sqrt{s}=X\,\mathrm{TeV}$}
\label{tab:numerical_results_analytical}
\end{center}
\end{table*}

\begin{figure}[h!]
	\begin{center}
		\includegraphics[scale=0.5]{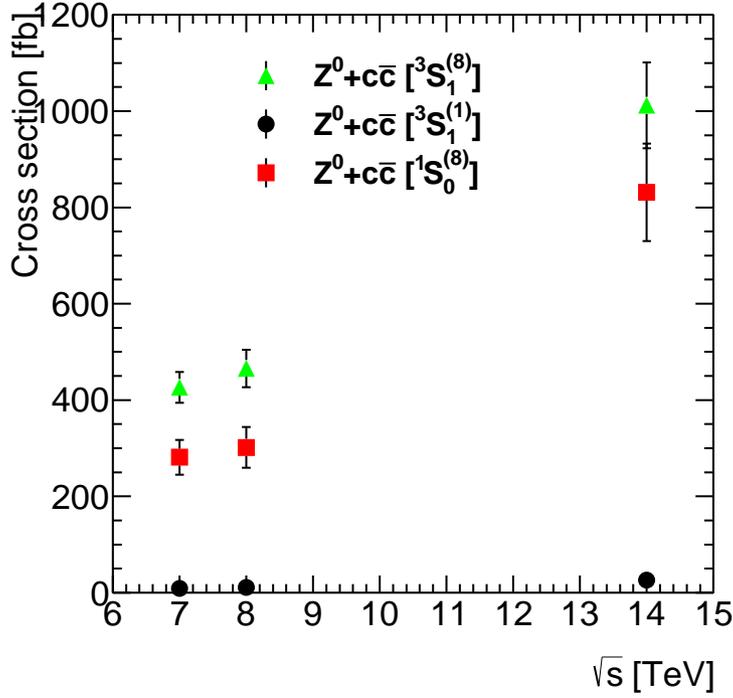}
		\caption{Cross section as a function of $\sqrt{s}$.}
 		\label{fig:results}
	\end{center}
\end{figure}

We calculated the cross section for the associated production of a $Z$ boson with a $J/\psi$ in proton-proton collisions to leading order. We have listed all the partonic contributions to the total cross section considering the $c\bar{c}[^{2S+1}L_J^{(c)}]$, with $S=1,2$, $L=S$, $J=0,1$ and $c=1,8$. The results obtained are visualized in figure \ref{fig:results} and summarized in table \ref{tab:numerical_results}.

The production of $Z$ in association with a $J/\psi$ is studied before in NLO accuracy \cite{ZJpsi2,ZJpsi3}. Based on our selections of $p_\mathrm{T}^{J/\psi}$ and renormalization and factorization scales we expect small next to leading order contributions.

\begin{table*}[h!]
\begin{center}\small
\begin{tabular}{ | c | c | c | c | c | c | }
  \hline
  & \multicolumn{3}{|c|}{cross section [$\mathrm{fb}$]} \\
  \hline
  Process & $\sqrt{s}=7\,\mathrm{TeV}$ & $\sqrt{s}=8\,\mathrm{TeV}$ & $\sqrt{s}=14\,\mathrm{TeV}$\\\hline\hline
  $Z + c\bar{c} [^3\mathrm{S}_1^{(8)}]$ & $ 426.6\pm 32.0$ & $465.8\pm 38.9 $ & $1012.1\pm   89.4 $\\\hline\hline
  $Z + c\bar{c} [^1\mathrm{S}_0^{(8)}]$ & $ 281.4\pm 36.2$ & $301.6\pm 42.5 $ & $831.3 \pm 101.3  $\\\hline\hline
  $Z + c\bar{c} [^3\mathrm{S}_1^{(1)}]$ & $ 8.8  \pm 1.0 $ & $11.1 \pm 1.3  $ & $25.9  \pm 3.3    $\\
  \hline
\end{tabular}
\caption{Cross sections tree level at $\sqrt{s}=X\,\mathrm{TeV}$}
\label{tab:numerical_results}
\end{center}
\end{table*}

It is clear that these processes are reachable within the statistics at the LHC, with the colour-octet process dominating. The observation of the associated production will provide a better determination of the $\langle\mathcal{O}^{J/\psi}[^{2S+1}L_J^{(c)}]\rangle$ elements and a good test of the NRQCD factorization formalism.

\section{Conclusions}

The vast statistics of the LHC can be proven useful in understanding and testing the quarkonia sector. Although the study of this sector begun in the late 70s, still there doesn't exist a model that can describe with high accuracy the experimental data. Associated production processes of electroweak boson and heavy quarkonia can be a powerful input for the models to produce more accurate predictions.

We studied the associated production of $Z+c\bar{c}\,(^{2S+1}L_J^{(c)})$, where $S=0,1$, $J=0,1$ and $c=1,8$. We find that the colour-octet process is dominantly contributing at the tree level. We listed all the parton contributions to the cross section of this process. We expect that with the collected luminosity at the LHC, there will be enough events to derive a cross section.

\subsection*{acknowledgments}
We are grateful to Costas Papadopoulos for the valuable comments.
\\
One of the authors (S.L.) wants to thank G. Karananas for the lengthy discussions.
\\
This research has been co-financed by the European Union (European Social Fund - ESF) and Greek national funds through the Operational Program "Education and Lifelong Learning" of the National Strategic Reference Framework (NSRF) - Research Funding Program: \textbf{THALES}. Investing in knowledge society through the European Social Fund.


\begin{thebibliography} {1}
\bibitem {quarkonium_group} {N. Brambilla, {\it et al.}, Eur. Phys. J. C 71, 1534, arXiv:1010.5827 (2011).}
\bibitem {ref:NRQCD} {S.M. Catterall {\it et al.}, Phys. Lett. B 300, 393 (1993).}
\bibitem {ref:CS1} {E.L. Berger {\it et al.}, Phys. Rev. D 23, 1521 (1981).}
\bibitem {ref:CS2} {R. Baier {\it et al.}, Phys. Lett. B102, 364 (1981).}
\bibitem{ref:Pwave} {A. Petrelli {\it et al.}, Nucl. Phys. B514, 245, arXiv:9707223 (1998).}
\bibitem{ref:Dwave} {Y. Fan {\it et al.}, Phys. Rev. D 80, 014001, arXiv:0903.4572 (2009).}
\bibitem{ref:CE1} {H. Fritzsch, Phys. Lett. 67B, 217 (1977).}
\bibitem{ref:CE2} {G. A. Schuler {\it et al.}, Phys. Lett. B387, 181 (1996).}
\bibitem{ref:NRQCD_lagrangian} {G. T. Bodwin {\it et al}, Phys. Rev. D 51, 1125 (1995).}
\bibitem {ref:fermiophobic} {M. A. Diaz, arXiv:9401259v1 (1994).}
\bibitem {ref:CDFZUpsilon} {D. Acosta, {\it et al.}, Phys. Rev. Lett., 90, 22 (2003).}
\bibitem{ref:ATLASWJpsi} {ATLAS collaboration, JHEP 04, 172 (2014).}
\bibitem{intrinsiccharm} {R. Vogt {\it et al.}, Phys. Lett. B349, 569-575, arXiv:9503206 (1995).}
\bibitem {MADONIA} {J. Alwall {\it et al.}, arXiv:1106.0522v1 (2011).}
\bibitem {QW1} {L. Gang {\it et al.}, JHEP01, 034 (2013).}
\bibitem {QW2} {E. Braaten {\it et al.}, Phys. Rev. D 60, 091501 (1999).}
\bibitem {QW3} {S. Mao {\it et al.}, arXiv:1304.4670 (2013).}
\bibitem {ZJpsi1} {B. A. Kniehl {\it et al.}, Phys. Rev. D 66, 114002 (2002).}
\bibitem {ZJpsi2} {B. Gong {\it et al.}, JHEP03,115 (2013).}
\bibitem {ZJpsi3} {S. Mao {\it et al.}, JHEP02, 071 (2011).}
\bibitem{ref:PDG} {Particle Data Group, J. Beringer {\it et al.}, Phys. Rev. D86, 010001 (2012).}
\bibitem{NRQCDitems} {Guang-Zhi Xu {\it et al.}, arXiv:1203.0207 (2013).}
\end{thebibliography}
\end{document}